\documentclass
[prb,10pt,a4paper,twocolumn,tightenlines,showpacs,preprint]{revtex4}%
\usepackage{amsmath}
\usepackage{graphicx}
\usepackage{amsfonts}
\usepackage{amssymb}%
\begin{document}
\title{Non-ohmicity and energy relaxation in diffusive 2D metals }
\author{Roy Ceder, Oded Agam, and Zvi Ovadyahu}
\affiliation{Racah Institute of Physics, The Hebrew University, Jerusalem 91904, Israel}

\begin{abstract}

We analyze current-voltage characteristics taken on Au-doped indium-oxide films.
By fitting a scaling function to the data, we extract the electron-phonon
scattering rate as function of temperature, which yields a quadratic dependence
of the electron-phonon scattering rate on temperature from 1K down to 0.28K. The
origin of this enhanced electron-phonon scattering rate is ascribed to the
mechanism proposed by Sergeev and Mitin.
\end{abstract}
\maketitle

\section{Introduction}

Energy relaxation processes play an important role in the low temperature
transport properties of diffusive metals, alloys and semiconductors. In
particular, they determine the maximum electric field $F$ allowed if the
system is to be measured under near-equilibrium conditions at a given
temperature. The condition for that may be expressed as $eFL_{\epsilon}\ll
k_{B}T$ where the energy relaxation length $L_{\epsilon}$ is the length over
which the electron diffuses under $F$ before the energy gained from it is
dissipated into the thermal bath. In particular, this issue is relevant for
all aspects of quantum transport such as corrections to the conductivity due
to interference and electron-electron interactions. For a system of size $L\gg
L_{\epsilon}$, energy relaxation processes are usually controlled by
electron-phonon inelastic scattering, and in the following we shall assume
that $L_{\epsilon}$ is dominated by the electron-phonon diffusion length
$L_{ep}$.

In clean samples the electron-phonon scattering mechanism is well understood,
and the scattering rate, $\tau_{ep}^{-1}$, is known to be proportional to
$T^{3}$ (where $T$ denotes the temperature). In dirty systems however, where
the elastic mean free path of the electron is smaller than the phonon thermal
length, the situation is more complicated. Schmid \cite{Schmid73} showed that
in this case $\tau_{ep}^{-1}$ is suppressed and becomes proportional to
$T^{4}$, in accordance with Pipard's ineffectiveness condition
\cite{Pippard55}. His model assumed that the impurities are anchored to the
lattice, and the scattering rate was calculated by moving into a reference
frame which follows the lattice vibrations. Riezer and Sergeev
obtained the same result using the laboratory frame of reference
\cite{Reizer86}. A scattering rate proportional to $T^{4}$ has indeed been
observed in a number of experiments \cite{Gershenson01,Karvonen04, Kivinen04}.
However, a $T^{3}$ law was frequently observed $\ $ even in systems that were
presumably well into the dirty limit regime \cite{Roukes85, Wennberg86,
Eshternach92, Wellstood94}. Moreover, quite a few observations of a $T^{2}$
scattering law were reported in other experiments \cite{Bergmann90,
DiTusa92,Watson95, Wu98}. The latter experimental results triggered further
studies with the aim of understanding better the electron-phonon scattering
mechanism in disordered metals. In particular, to obtain an electron-phonon
scattering rate proportional to $T^{2}$ (rather than the "ineffective" $T^{4}$
law), Belitz and Wybourne \cite{Belitz95} assumed a strong phonon damping,
while Jan Wu and Wei \cite{Jan05} included effects associated with the
discrete lattice structure. Sergeev and Mitin \cite{Sergeev00} obtained the
$T^{2}$ behavior from a model were impurities are assumed to be fixed, namely,
impurities which do not follow the lattice vibrations. They argued that heavy
impurities or boundaries which move differently from the host lattice produce
the same effect.

\begin{figure*}
\includegraphics[width=79mm]{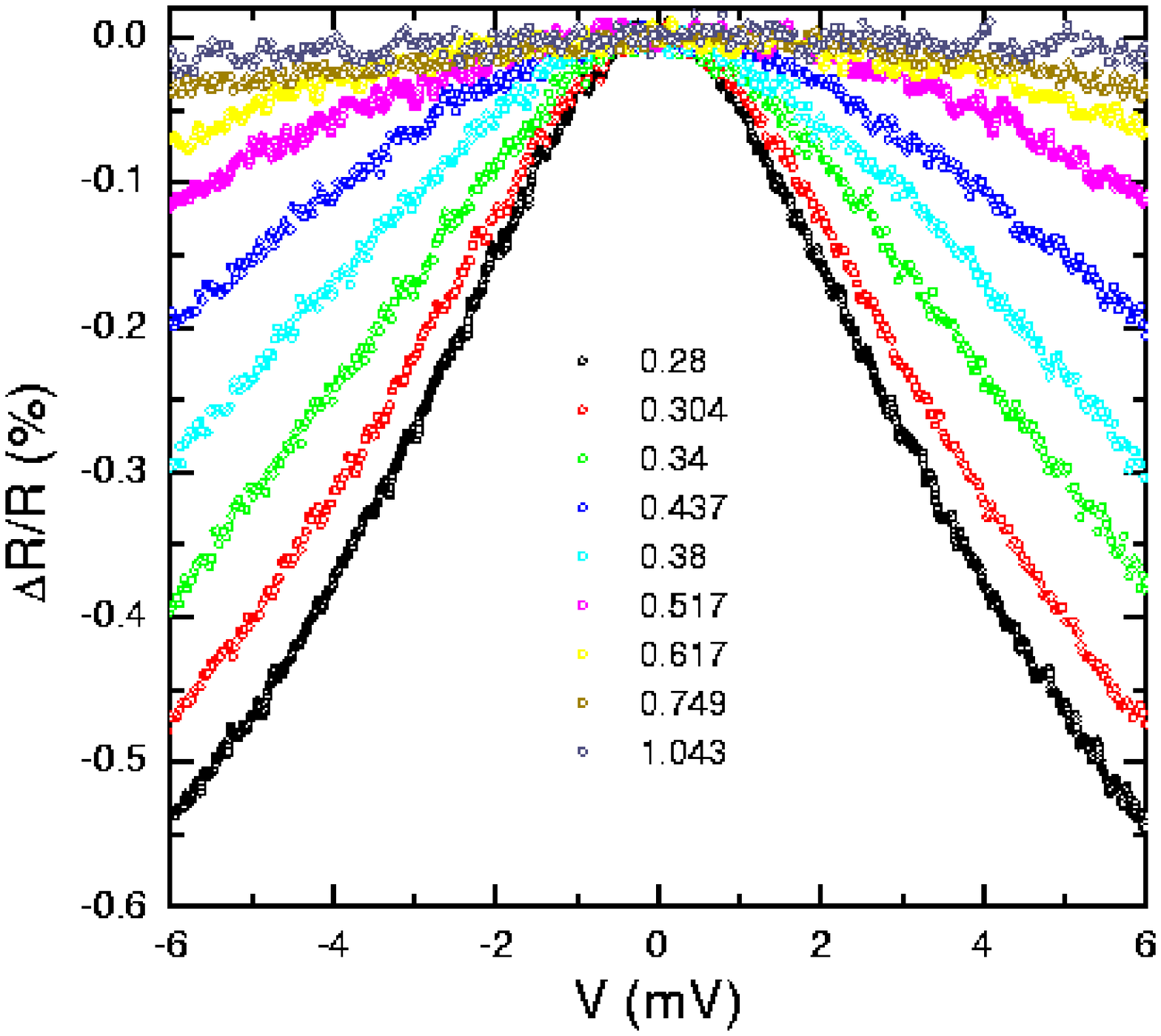}
\includegraphics[width=8.1cm]{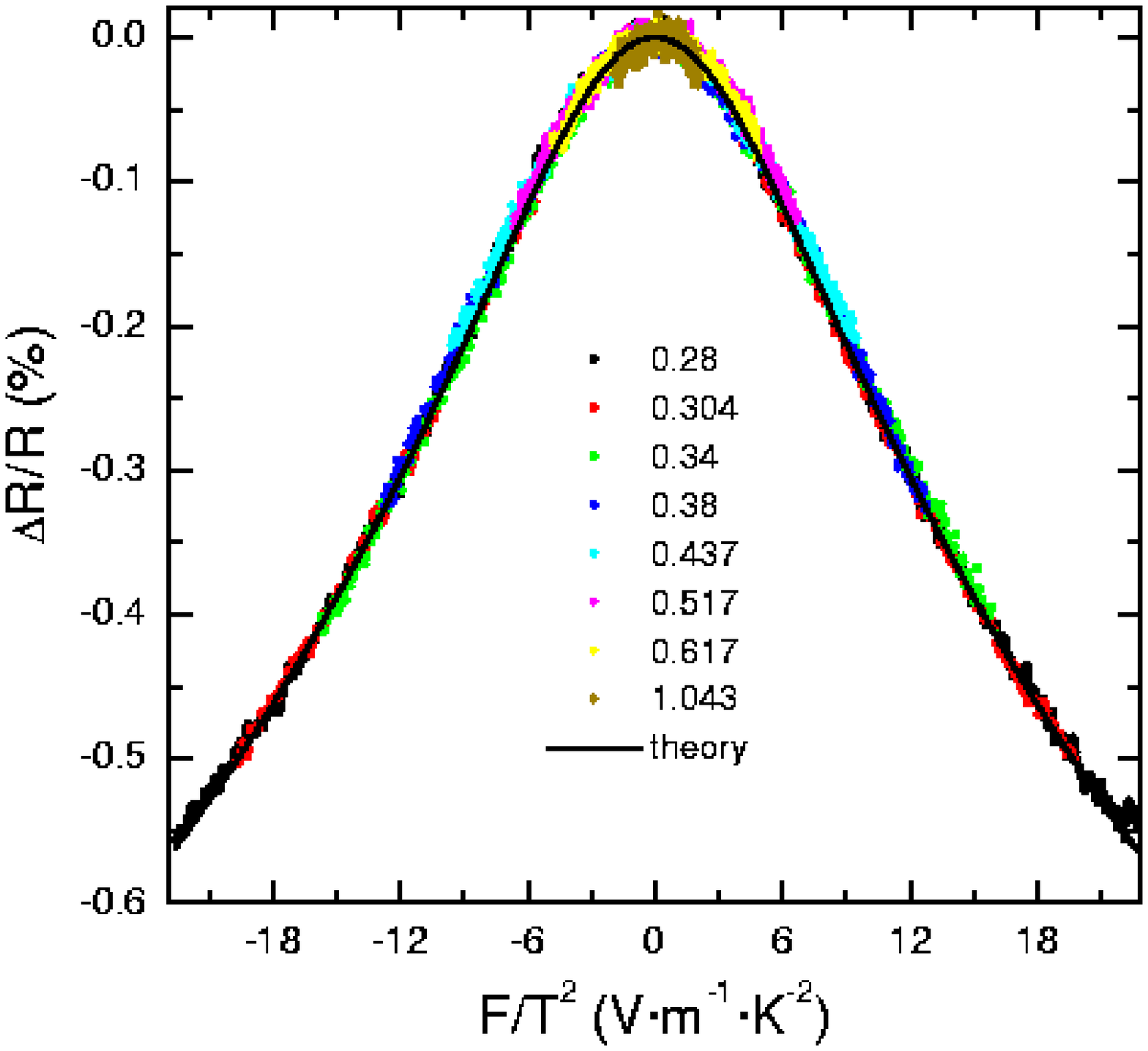}
\caption{ The non-ohmic characteristic of 200\AA  thick In$_{2}$O$_{3-x}$:Au sample with
length L=3500$\mu m$ and width W=1mm. Left panel: The differential resistance
(defined in Eq.~(\ref{scaling-function}) as function of the voltage. Right
panel: The same data plotted as function of $F(=V/L)$ normalized by $T^{2}$.
Note the near perfect data collapse as well as the fit to formula
(\ref{R-scaling-1}) represented by the continuous line.}%
\end{figure*}

In this paper we analyze the non-ohmic characteristic of thin films of
In$_{2}$O$_{3-x}$:Au (crystalline indium-oxide doped with 2\% gold), that were
characterized and measured as described elsewhere \cite{Zvi_1}. We show that
in this system $\tau_{ep}^{-1}\propto T^{2}$, and interpret this behavior as a
manifestation of the Sergeev Mitin mechanism where the Au atoms play the role
of the \textquotedblleft immobile impurities\textquotedblright.  The sample
resistance is used as a thermometer of the electron temperature. The latter is determined
 by the energy balance between the Joule heating, due to the presence of electric field, 
and the heat transfer to the lattice phonons governed by $\tau_{ep}^{-1}.$

\bigskip The data we shall analyze are dynamic resistance $R$ versus voltage
traces for a typical film at $T\leq1K$ shown in Fig.~1 (left panel). As
expected, the deviations from ohm's law become more pronounced as $T$
decreases. In the right panel of Fig.~1 it is shown that these curves taken at
different temperatures can be collapsed onto a common function:
\begin{equation}
\frac{\Delta R}{R}\equiv\frac{R(F,T)-R(0,T)}{R(0,T)}=\Delta\mathcal{R}%
_{p}\left(  \frac{F}{T^{\frac{p}{2}+1}}\right)  . \label{scaling-function}%
\end{equation}
with $p=2$. The scaling of the electric field as a power of the temperature,
$F\sim T^{\frac{p}{2}+1}$, and its relation to the electron-phonon time
$\tau_{ep}^{-1}(T)\propto T^{p}$, has been already recognized by Anderson,
Abrahams and Ramakrishnan \cite{Anderson79}, and later by Arai \cite{Arai83}.
In the next section we shall calculate the function $\Delta\mathcal{R}_{p}$,
 and clarify its relation to the electron-phonon relaxation time. We shall 
consider the dependence on the sample length and
compare with further experimental results in section 3.

\section{Theory}

\bigskip The essence of the picture developed below is the assignment of an
effective temperature for a given field and bath-temperature. The electrons
are accelerated by the electric field, and the collisions with other electrons
and phonons result in a local equilibrium distribution characterized by an
effective temperature $T_{eff}$. The latter differs from the bath temperature
$T$, and depends on the electric field. We shall make the association
$R(F,T)\simeq R(0,T_{eff})$. Hence knowing $T_{eff}(F,T)$ and the form of the
near-equilibrium $R(0,T)$ yields the desired $R(F,T)$ from which we deduce the
scaling function (\ref{scaling-function}).

We begin by considering the Boltzmann equation for the electron distribution
function $f$:
\begin{equation}
\frac{\partial f}{\partial t}+\vec{v}\cdot\frac{\partial f}{\partial\vec{r}%
}+e\vec{F}\cdot\frac{\partial f}{\partial\vec{p}}=I[f]. \label{boltzmann}%
\end{equation}
Here $\vec{r}$ is the position, $\vec{v}$ is the velocity, $\vec{p}$ denotes
the momentum, $\vec{F}$ the electric field, and $I[f]=I_{im}[f]+I_{ee}%
[f]+I_{ep}[f,N]$ represents the collision integrals due to impurity
scattering, electron-electron, and electron-phonon interactions. The latter
depends also on the phonon distribution function, assumed to be the
equilibrium distribution function denoted by $N$.

Following Nagaev\cite{Nag_1} and Kozub and Rudin\cite{Kozub}, we look for a
steady state solution of the form:%
\begin{equation}
f=f(\vec{r},\hat{n},\epsilon-e\vec{F}\cdot\vec{r}) \label{f-form}%
\end{equation}
where $\hat{n}$ denotes a unit vector in the direction of the momentum, and
$\epsilon=\epsilon(p)$ is the energy assumed to depend on the absolute value
of the momentum $p=|\vec{p}|$. Substituting (\ref{f-form}) in (\ref{boltzmann}%
) leads to:
\begin{equation}
\vec{v}\cdot\frac{\partial f}{\partial\vec{r}}+\sum_{ij}\frac{eF_{i}}%
{p}\left(  \delta_{ij}-n_{i}n_{j}\right)  \frac{\partial f}{\partial n_{j}%
}=I[f]. \label{boltzmann1}%
\end{equation}
Next, we define the symmetric and anti-symmetric parts of the distribution
function with respect to the momentum direction $f^{\pm}=\left(  f(\hat{n})\pm
f(-\hat{n})\right)  /2$, and construct two new equations from
(\ref{boltzmann1}) associated with the addition and subtraction of Boltzmann
equations for $f(\hat{n})$ and $f(-\hat{n})$. Then assuming that momentum
relaxation is dominated by scattering from impurities, and that
electron-electron and electron-phonon interactions are essentially independent
of the momentum direction (i.e., $I_{ee}[f]\simeq I_{ee}[f^{+}]$, and
$I_{ep}[f,N]\simeq I_{ep}[f^{+},N]$), we obtain:
\begin{equation}
\vec{v}\frac{\partial f^{-}}{\partial\vec{r}}+\sum_{ij}\frac{eF_{i}}{p}\left(
\delta_{ij}-n_{i}n_{j}\right)  \frac{\partial f^{-}}{\partial n_{j}}=\bar
{I}[f^+], \label{boltzmann+}%
\end{equation}
where $\bar{I}[f^+]\simeq I_{ee}[f^{+}]+I_{ep}[f^{+},N]$, and
\begin{equation}
\vec{v}\frac{\partial f^{+}}{\partial \vec{r}}=I_{im}[f^{-}]. \label{boltzmann-}%
\end{equation}

In the simplest approximation, the impurity collision term takes the form
$I_{im}[f^{-}]=-f^{-}/\tau$ where $\tau$ is the elastic mean free time. Thus
using (\ref{boltzmann-}), the antisymmetric part of the distribution function
can be expressed in terms of the symmetric part and substituted into
(\ref{boltzmann+}). The resulting equation is now averaged over the momentum
directions to give
\begin{equation}
-D\nabla^{2}f^{+}=I_{ee}[f^{+}]+I_{ep}[f^{+},N], \label{diffuion-boltzmann}%
\end{equation}
where $D=\tau v_{F}^{2}/3$ is the diffusion constant ($v_{F}$ is the Fermi
velocity). Here and henceforth we neglect the energy dependence of the
diffusion constant.

We wish to solve Eq.~(\ref{diffuion-boltzmann}) for homogeneous quasi-two
dimensional samples of rectangular shape with voltage contacts located at
$x=\pm L/2$. Thus $f^{+}$ is independent of the transverse coordinate, and the
boundary conditions assuming ideal contacts are:
\begin{equation}
\left.  f^{+}\right\vert _{x=\pm L/2}=n_{F}\left(  \frac{\epsilon-\epsilon
_{F}\mp\frac{eV}{2}}{k_{B}T}\right)  , \label{f-boundary-conditions}%
\end{equation}
where $n_{F}(x)=(1+\exp(x))^{-1}$ is the Fermi distribution function,
$\epsilon$ is the electron energy, $\epsilon_{F}$ is the Fermi energy, $V=FL$
is the voltage drop across the sample, $k_{B}$ is Boltzmann constant, and $T$
is the bath temperature.

Equation (\ref{diffuion-boltzmann}) is a nonlinear equation for the electron
distribution function. To make progress, we shall assume that the
electron-electron diffusion length $L_{ee}$ is much smaller than the system
size, $L$, and the energy relaxation length $L_{ep}$. This should secure an
effective local thermalization due to the large number of collisions an
electron experiences. Looking then for a solution which describes local
equilibrium of the electrons:
\begin{equation}
f^{+}=n_{F}\left(  \frac{\epsilon-\epsilon_{F}-eFx}{k_{B}T_{loc}(x)}\right)  ,
\label{anstz}%
\end{equation}
where $T_{loc}(x)$ is a local temperature of the electrons. At the contacts
the local temperature by assumption coincides with the bath temperature:
\begin{equation}
T_{loc}\left(  \pm\frac{L}{2}\right)  =T, \label{boundaryT}%
\end{equation}
so that the solution (\ref{anstz}) with the boundary conditions
(\ref{boundaryT}) satisfies the requirement (\ref{f-boundary-conditions}).
Under the assumption of local equilibrium the electron-electron collision term
vanishes and equation (\ref{diffuion-boltzmann}) reduces to
\begin{equation}
-D\frac{\partial^{2}f^{+}}{\partial x^{2}}=I_{ep}[f^{+},N].
\end{equation}
Finally, to extract the local temperature behavior we multiply this equation
by $\epsilon$ and integrate over the energy. The resulting equation is:
\begin{equation}
D\left[  \frac{\pi^{2}k_{B}^{2}}{6}\frac{\partial T_{loc}^{2}(x)}{\partial
x^{2}}+(eF)^{2}\right]  =-\int  d\epsilon \epsilon I_{ep}[f^{+},N].
\label{local-temperature}%
\end{equation}
To further simplify (\ref{local-temperature}), we take the electron-phonon 
collision integral to have
the form:
\begin{align}
I_{ep}  &  =\int d\omega K(\omega)\left[  -\left(  1-f^{+}(\epsilon
+\omega)\right)  f^{+}(\epsilon)N\left(  \frac{\omega}{k_{B}T}\right)  \right.
\nonumber\\
&  +\left(  1-f^{+}(\epsilon)\right)  f^{+}(\epsilon+\omega)\left(  N\left(
\frac{\omega}{k_{B}T}\right)  +1\right) \nonumber\\
&  -\left(  1-f^{+}(\epsilon-\omega)\right)  f^{+}(\epsilon)\left(  N\left(
\frac{\omega}{k_{B}T}\right)  +1\right) \nonumber\\
&  \left.  +\left(  1-f^{+}(\epsilon)\right)  f^{+}(\epsilon-\omega)N\left(
\frac{\omega}{k_{B}T}\right)  \right]  , ~~~~~~~~ \label{ep-collision}%
\end{align}
where $K(\omega)=\alpha\omega^{p-1}$, is a kernel depending on the nature of
the collision between the electrons and phonons, while $N(x)=(\exp(x)-1)^{-1}$
is the equilibrium phonon  distribution function. We substitute this expression
into (\ref{local-temperature}) with the approximate local equilibrium form of
the electron distribution function (\ref{anstz}), and integrate over $\epsilon$
and $\omega$. The resulting equation is an equation for the local temperature
\begin{equation}
\frac{\pi^{2}k_{B}^{2}}{6}\frac{\partial T_{loc}^{2}(x)}{\partial x^{2}%
}+(eF)^{2}=\eta k_{B}^{p+2}\left[  T_{loc}^{p+2}(x)-T^{p+2}\right]  ,
\label{local-temp-eq}%
\end{equation}
where
\begin{equation}
\eta=(1-2^{-(p+1)})(p+1)!\zeta(p+2)\frac{\alpha}{D}.
\end{equation}

In understanding the form of the solution of equation~(\ref{local-temp-eq}),
it is instructive to identify first the relevant length scale for this
equation. To this end one may linearize the equation by substituting
$T_{loc}^{2}(x)=T^{2}+\delta T^{2}(x)$ and expanding to linear order in
$\delta T^{2}(x)$. The solution of the resulting equation with the boundary
conditions (\ref{boundaryT}) is
\begin{equation}
T_{loc}^{2}(x)=T^{2}+\frac{6(eFL_{ep})^{2}}{k_{B}^{2}\pi^{2}}\left(
1-\frac{\cosh\left(  \frac{x}{L_{ep}}\right)  }{\cosh\left(  \frac{L}{2L_{ep}%
}\right)  }\right)  , \label{linear-solution}%
\end{equation}
where
\begin{equation}
L_{ep}=\frac{\pi\left(  k_{B}T\right)  ^{-\frac{p}{2}}}{\sqrt{(3(p+2)\eta}}
\label{ep-length}%
\end{equation}
is essentially the electron-phonon length at equilibrium. From
(\ref{linear-solution}) one can see that $L_{ep}$ sets the distance from the
contacts over which the temperature profile reaches a saturated value. Thus
long sample satisfies $L\gg L_{ep}$, and these may be considered to have an
essentially space independent local temperature which we shall refer to as the
effective temperature.

The linearized solution for long samples (\ref{linear-solution}) is strictly
justified when the electric field is weak, i.e. $eFL_{ep}<k_{B}T$. We shall be
interested in the limit of strong fields where $L_{ep}$ will presumably be
smaller than its equilibrium value, the r.h.s of (\ref{ep-length}). Provided
$L\gg L_{ep}(T,F)$ it makes sense to assume an electron temperature, which is
essentially constant throughout the sample. We then neglect the space
dependent term in equation~(\ref{local-temp-eq}), and the effective
temperature of the electrons, at any field strength, is readily deduced to be:
\begin{equation}
T_{eff}\simeq\left[  T^{p+2}+\frac{(eF)^{2}}{\eta k_{B}^{p+2}}\right]
^{\frac{1}{p+2}}.
\end{equation}

As mentioned earlier, once $T_{eff}$ is known, to find the scaling function
$\Delta\mathcal{R}_{p}$ requires only the temperature dependence of the
resistance, $R(T)$, since $R(F,T)\simeq R(0,T_{eff})$. At low temperatures,
the temperature dependent terms of the resistance are the weak localization
\cite{Brg_1} and the Altshuler-Aronov corrections \cite{alt_1}. For thin
films, both of these have a logarithmic behavior, thus
\begin{equation}
R(0,T)\simeq R_{D}(1-\gamma\ln(T)), \label{equilibrium-resistance}%
\end{equation}
where $\gamma$ is a small dimensionless constant, depending on the nature of
the electron-electron interactions, the spin-orbit coupling, and the ratio of
the quantum unit resistance to the Drude resistance of the sample, $R_{D}$.
Substituting $R(F,T)\simeq R(0,T_{eff})$ and (\ref{equilibrium-resistance}) in
the definition of the scaling function (\ref{scaling-function}), and expanding
to the leading order in $\gamma$ we obtain:
\begin{equation}
\Delta\mathcal{R}_{p}\left(  \frac{F}{T^{\frac{p}{2}+1}}\right)  \simeq
-\frac{\gamma}{p+2}\ln\left[  1+\frac{(eF)^{2}}{\eta\left(  k_{B}T\right)
^{p+2}}\right]  , \label{R-scaling-1}%
\end{equation}
Note that this has the scaling form as in Eq.~(\ref{scaling-function}) above.

To anticipate the discussion in the next section, it is instructive to
consider the case of short samples $L_{ep}\gg L$. Here one may neglect the
contribution of the electron-phonon collision term in Eq.~(\ref{local-temperature}) 
and readily obtain the solution for
$T_{loc}(x)$, whose space dependence, in general, cannot be ignored:
\begin{equation}
T_{loc}^{2}(x)=T^{2}+\frac{3(eF)^{2}}{k_{B}^{2}\pi^{2}}\left(  \frac{L^{2}}%
{4}-x^{2}\right)  . \label{local_T}%
\end{equation}
At the temperature range where the dephasing length is much smaller than the
system size one may view the sample as a set of classical resistors connected in
series. Therefore the total resistance can be approximated by the sum:
$R(F,T)\simeq\sum_{j}R_{j}$, where $R_{j}=R\left(  T_{loc}(x_{j})\right)  $ is
the resistance of the j-th segment (of size of the dephasing length), centered at the point
$x_{j}$. Thus, assuming homogeneous sample, the experimentally measured
resistivity is essentially an average over the position. From this average one
immediately obtains the scaling function of short samples:
\begin{equation}
\Delta\mathcal{R}_{0}\left(  \frac{V}{T}\right)  \simeq-\gamma\left[
\chi\tanh^{-1}\left(  \frac{1}{\chi}\right)  -1\right]  ,
\label{short-scaling}%
\end{equation}
where
\begin{equation}
\chi=\sqrt{1+\frac{4\pi^{2}}{3}\left(  \frac{k_{B}T}{eV}\right)  ^{2}},
\end{equation}
and $V=FL$ is the voltage drop along the sample.

\section{Data analysis and discussion}

Comparing the resistance curves shown in Fig.~1 with Eq.~(\ref{R-scaling-1}),
one notes that a scaling form that leads to the data collapse occurs for
$p=2$. This means that the energy relaxation time is quadratic with
temperature: $\tau_{ep}^{-1}\propto\frac{D}{L_{ep}^{2}}\propto T^{2}$, and
therefore the electron-phonon length (\ref{ep-length}) is inversely
proportional to the temperature.

\begin{figure}[ptb]
\includegraphics[width=8cm]{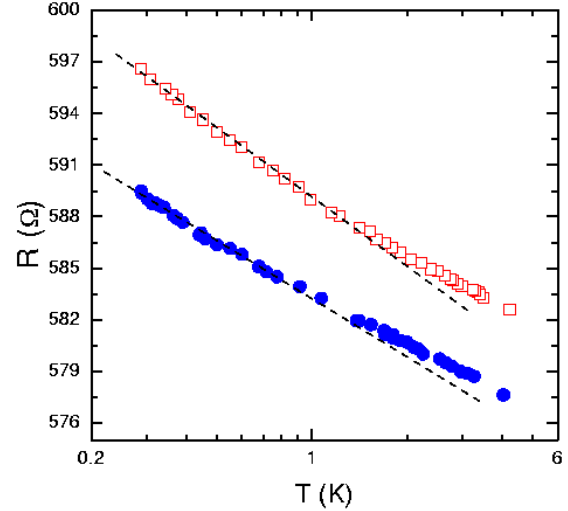}
\caption{Resistance versus temperature for the sample of Fig.~1 (open
squares) and for a In$_{2}$O$_{3-x}$:Au sample with length L=80$\mu m$ and
width W=500$\mu m$ (full circles). The dashed lines are the logarithmic slopes
used in defining the value of $\gamma$ (see Eq.~(\ref{equilibrium-resistance})).}%
\end{figure}

The procedure of extracting the detailed form of $\tau_{ep}^{-1}$ or $L_{ep}$
from the experimental results is as follows. First, the value of $\eta$ is
obtained by fitting the $R(F,T)$ data of Fig.~1 to Eq.~(\ref{R-scaling-1}),
 as shown in the right panel of this figure. Note that the latter needs the
parameter $\gamma$ as input. This is defined by equation
(\ref{equilibrium-resistance}) above and is thus obtained from the
near-equilibrium $R(T)$ measurement performed on the same sample. Such $R(T)$
data and their associated $\gamma$ are shown in Fig.~2 for two samples. These
are made from the same batch of a Au-doped $In_{2}O_{3-x}$ film, they only
differ in their lateral dimensions. The first is the 3500$\mu$m sample of
Fig.~1. The second sample is $80\mu m$ long. The sheet resistances of the two
are within 1\% of each other, yet their logarithmic slopes are somewhat
different. Also, both samples show a systematic deviation from the theoretical
$\ln(T)$ dependence, a feature that seem to occur in some other 2D systems
\cite{Mueller94}. Since we are mainly interested in the restricted temperature
range $0.28-1K$, this feature may be ignored, and $\gamma$ is defined by
fitting $R(T)$ to a simple $\ln(T)$ over the relevant range as shown in the
figure. The fits yield $\gamma\approxeq0.0098$ and $\gamma\approxeq0.0081$ for
the long and short samples respectively, and these values are used in the
subsequent analysis below. 
\begin{figure}[ptb]
\includegraphics[ width=8cm]{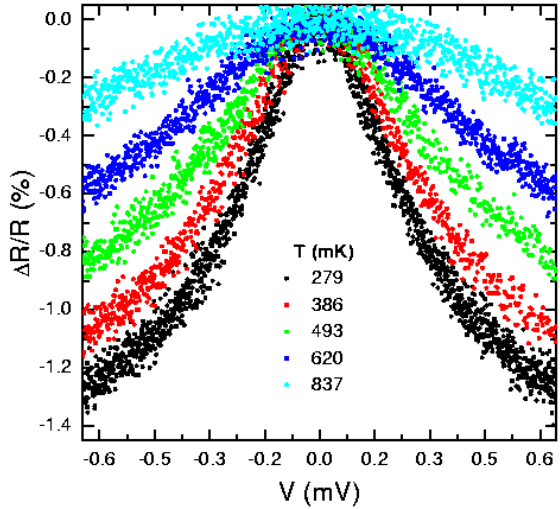}
\caption{Dependence of the differential resistance on voltage at several
temperatures. In$_{2}$O$_{3-x}$:Au sample with length L=80$\mu m$ and width
W=500$\mu m$. The noisier data here (as compared with those of Fig.~1), is
mainly due to the much smaller sample size.}%
\end{figure}

\begin{figure*}
\includegraphics[width=8cm]{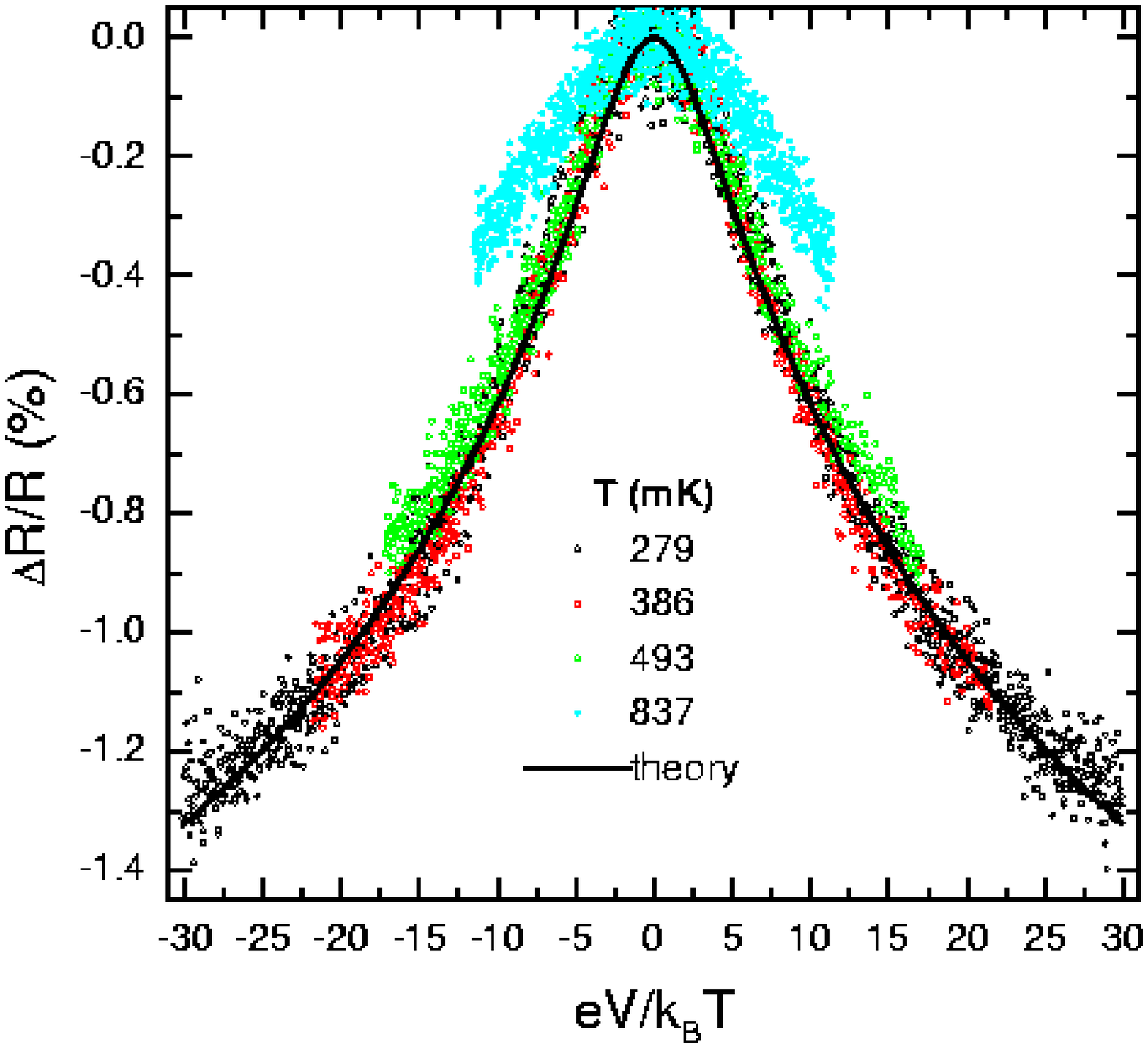}
\includegraphics[width=8.2cm]{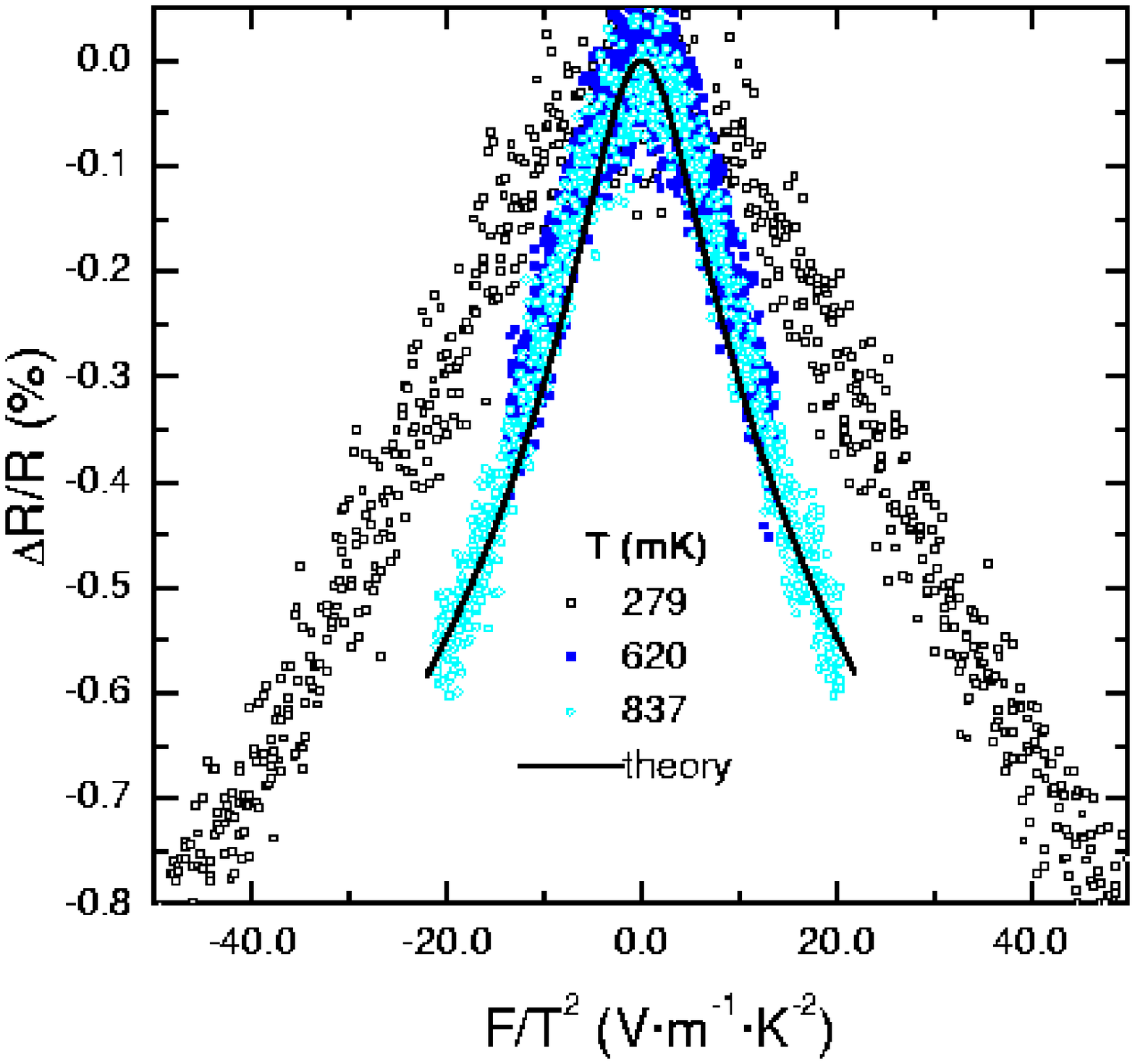}
\caption{Left panel: The differential resistance versus the voltage $V$
normalized by the temperature $T$ using the data of Fig.~3. The full line is a
best fit to the data using Eq.~(\ref{short-scaling}) . Note the data collapse
for the three lowest temperatures and the deviations from the short-sample
behavior at $T=837mK$.Right panel: Data from the same figure (Fig.~3) plotted
as function of $F(=V/L)$ normalized by $T^{2}$. The full line is a best fit to
the data using Eq.~(\ref{R-scaling-1}). Note the data collapse for the two
highest temperatures and the deviations from the long-sample behavior at
$T=279mK$.}%
\end{figure*}

An excellent fit to the data in the right panel of Fig.~1 can be
obtained using Eq.~(\ref{R-scaling-1}) with $\eta = 3.7\cdot
10^{55}\frac{1}{Joule^{2}m^{2}}$, $\gamma= 0.0098$ and $p=2$. The use
of this formula, appropriate for the $L\gg L_{ep}$ limit, is justified for
this sample as can be seen by estimating $L_{ep}$. Inserting the above value
of $\eta$ in equation (\ref{ep-length}), gives $L_{ep}\simeq$ $20\mu m$ at
$1K$ and only $L_{ep}\simeq60\mu m$ at $T\simeq 0.28K$. Thus $L_{ep}$ is much
smaller than $L$ down to the lowest temperature we are dealing with here.

For the 80$\mu m$ sample, on the other hand, a crossover to the short-sample
regime is realized in the temperature interval covered in our experiments. The
crossover can be seen by studying the data depicted in Figs. 3, and 4. Fig. 3
shows the raw $R(V)$ data measured at different temperatures. Fig.~4 show
these data plotted according to the short-sample formula
(Eq.~(\ref{short-scaling})) in the left panel,  and  according to the long-sample scheme
(Eq.~(\ref{R-scaling-1}) with $p=2$) in the right panel. The crossover temperature is the
temperature below which the $R(F,T)$ data can be scaled as a function of
$V/T$, and above which it scales as $F/T^{2}$. Comparing between the left and
the right panels of Fig.~4 it is evident that this temperature is
approximately $0.5K$. Thus, with this sample, the consistency of our 
approach can be tested in the two limits. As shown in the left panel of Fig.~4,
a good fit to the $\Delta R/R(0)$ data is obtained using
Eq.~(\ref{short-scaling}), which involves just the parameter $\gamma$. Note
that the best fit $\gamma$ is quite close to the $\gamma$ that one gets from
the logarithmic fit to the $R(T)$ data of this sample (Fig.~2). The other
limit, which conforms to Eq.~(\ref{R-scaling-1}), also yields a reasonable
agreement, with the \textit{same} $\gamma$ as used above, and with
$\eta=2.2\cdot10^{55}\frac{1}{Joule^{2}m^{2}}$ (see right panel of Fig.~4).
The quality of the fit here is less good than in the 3500$\mu m$ sample,
perhaps due to the fact that even at the highest temperature used the sample
is not really in the long-sample limit.

Finally, using the value of $\eta$ for this sample in Eq.~(\ref{ep-length}),
one gets $L_{ep}\simeq$ $34\mu m$ at $1K$ and$\simeq100\mu m$ at
$T\approxeq0.28K$ . Since it is plausible to expect that the crossover from
the short-sample to long-sample regime should occur when $L_{ep}$ becomes
comparable with $L/2$, these numbers are consistent with our picture.

We turn now to discuss the physics that underlies the $\tau_{ep}^{-1}\propto
T^{2}$ law for the electron-phonon scattering rate suggested by our analysis.
It is important to note that the enhanced electron-phonon inelastic scattering
resulted from the inclusions of Au atoms in the indium-oxide matrix: By
comparison, an undoped In$_{2}$O$_{3-x}$ sample showed $\tau_{ep}^{-1}$ that,
at $T\approx0.5K,$ was more than \textit{three orders of magnitude smaller}
\cite{Zvi_1}. Since these 'pure' and the Au-doped samples had otherwise quite
similar parameters (their $R_{\square},$ and diffusion constant were the same
to within 30\% ), the non-trivial role of the gold in enhancing $\tau
_{ep}^{-1}$\ must be considered. It seems likely that this is a manifestation
of the Sergeev-Mitin mechanism for electron-phonon scattering in disordered
metals. The gold impurities in our samples are heavier than the host atoms,
and being inert they are also loosely attached to the indium-oxide lattice.
These factors limit their ability to follow the lattice movement, and thus the
main assumption of the Sergeev-Mitin mechanism
is fulfilled. At the same time, the Au atoms are active as local soft-modes
which could be very effective in dephasing the electrons\cite{Imry}.
However, being weakly coupled to the lattice, they cannot efficiently dissipate the
energy, gained by inelastic collisions with the electrons,
to the bath.  Therefore the Au inclusions contribute to dephasing 
much more than to energy relaxation. Note indeed that the phase coherence length in these 
samples  is dominated by the interaction of the electrons with these local modes\cite{Zvi_1}
and it is $\approx0.4\mu m$ at $T=0.3K$ as compared with $L_{ep}\approx$ $100\mu m$.
That the dephasing rate exceeds the energy relaxation rate by many orders of
magnitude is quite a general property of low temperature transport, which
follows from the different$\ $temperature dependencies of energy relaxation
processes on one hand and dephasing on the other hand.

To summarize, we have employed a scaling analysis of non-ohmic resistance
curves in order to extract the electron-phonon scattering rate of metallic
films. The method makes use of the temperature dependence of the resistivity
therefore it is best suited for those cases where the resistance can be used
as a sensitive thermometer. Our analysis, which assumes quasi-two dimensional
samples may be easily extended to other dimensionalities. In these cases the
temperature dependence of the resistance, at sufficiently low temperatures, is
dominated by a power low behavior, $R(T)=R_{D}(1-\gamma T^{\nu})$. From here
it follows that the scaling function (\ref{scaling-function}) of long samples
satisfies the relation:
\[
T^{-\nu}\Delta\mathcal{R}=\gamma\left[  1-\left(  1+\frac{(eF)^{2}}{\eta
T^{p+2}}\right)  ^{\frac{\nu}{p+2}}\right]  .
\]
Having the equilibrium parameters ($\gamma$ and $\nu$), the experimental data
of $R(T,F)$ can be fitted to the above form and both $\eta$ and $p$ can be
extracted. The electron phonon length is then deduced from Eq.~(\ref{ep-length}).

We reiterate that in order to apply the scaling approach, the following
conditions should be satisfied: (a) The heat transfer from the electrons to
the bath is dominated by the electron-phonon collisions; (b) the
electron-electron diffusion length should be much smaller than the energy
relaxation length.

On the other hand, the scaling approach is insensitive to the inclusion of
other ingredients such as two level systems and Kondo impurities as long as
they do not serve as additional channels for heat conduction to the bath.
Furthermore, for long samples the quality of the contacts is of minor
importance, since the amount of heat transferred by the electrons through the
contacts is anyhow negligible. For short samples, however, our analysis
assumes that the electrons near the contacts are at the bath temperature. This
means that the contacts are ideal heat sinks, a caveat that should be borne in
mind when using contacts made of a superconducting material. \bigskip\ 

The authors gratefully acknowledge discussions with I.~Aleiner and Y.~Imry.
This work has been supported in part by the Israel Science Foundation (ISF), 
and by the German-Israel Foundation (GIF).

\end{document}